\documentclass[prd,twocolumn,nofootinbib,reprint]{revtex4-1}
\usepackage{amsmath,amssymb,amsthm,bm,braket,ascmac,url,hyperref,color,comment} 
\usepackage{graphicx}
\begin{document}
\title{Absolute Lower Bound on the Bounce Action}
\author{Ryosuke Sato$^1$ and Masahiro Takimoto$^{1,2}$}
\affiliation{\vspace{2mm} $^1$Department of Particle Physics and Astrophysics, Weizmann Institute of Science, Rehovot 7610001, Israel \\
$^2$Institute of Particle and Nuclear Studies, High Energy Accelerator Research Organization (KEK),\\
Tsukuba 305-0801, Japan}

\date{\today}

\begin{abstract}
\vspace{1mm}
The decay rate of a false vacuum is determined by the minimal action solution of the tunnelling field: \textit{bounce}.
In this Letter, we focus on models with scalar fields which have a canonical kinetic term in $N(>2)$ dimensional Euclidean space,
and derive an absolute lower bound on the bounce action.
In the case of four-dimensional space,
we show the bounce action is generically larger than $24/\lambda_{\rm cr}$,
where $\lambda_{\rm cr} \equiv {\rm max} [ -4V(\phi) /|\phi|^4] $  with the false vacuum being at $\phi=0$ and $V(0)=0$.
We derive this bound on the bounce action \textit{without solving the equation of motion explicitly}.
Our bound is derived by a quite simple discussion,
and it provides useful information even if it is difficult to obtain the explicit form of the bounce solution.
Our bound offers a sufficient condition for the stability of a false vacuum,
and it is useful as a quick check on the vacuum stability for given models.
Our bound can be applied to a broad class of scalar potential with any number of scalar fields.
We also discuss a necessary condition for the bounce action taking a value close to this lower bound.
\end{abstract}
\maketitle

\section{Introduction}\label{sec:intro}
The stability condition of a vacuum is one of the important constraints on viable models of particle physics.
Even in the standard model, it gives a nontrivial constraint on the Higgs boson mass and the top quark mass \cite{Degrassi:2012ry}.
Furthermore, physics beyond the standard model often introduces additional scalar fields,
and they could destabilize the standard model vacuum by giving a deeper vacuum.
In these situations, the standard model vacuum is a false vacuum and its lifetime should be longer than the age of the Universe.

The lifetime of a false vacuum in quantum field theory can be calculated by using Coleman's semiclassical method \cite{Coleman:1977py}.
In this method, the decay rate of a false vacuum per volume is evaluated as $\Gamma/V \sim A e^{-S}$
where $A$ is a prefactor and $S$ is the action for a nontrivial solution of the equation of motion which gives the minimal action.
Such a solution is called a bounce solution.
To obtain the bounce solution,
we have to solve the equation of motion of scalar fields with an appropriate boundary condition.
However, it is not always easy to obtain the explicit solution of the equation of motion.
In particular, we have to solve a large number of coupled equations of motion
if we consider some model with a large number of scalar fields such as the landscape scenario \cite{Greene:2013ida, Dine:2015ioa}.

It is convenient if we can discuss a possible range of the minimal bounce action value without solving the equation of motion explicitly.
In this context, for example, a generic upper bound on the minimal bounce action is discussed in Refs.~\cite{Dasgupta:1996qu, Sarid:1998sn}.
A lower bound is discussed in Ref.~\cite{Aravind:2014aza} which focuses on quartic scalar potential,
and Ref.~\cite{Masoumi:2017trx} which reduces the problem to the effective single scalar problem.

In this Letter, we derive a generic lower bound on the minimal bounce action,
which can be applied to a broad class of scalar potential with any number of scalar fields.
Our bound can be derived by using a quite simple discussion which is based on the Lagrange multiplier method.
The bound has a simple form, and it provides a sufficient condition for the stability of a false vacuum.
Therefore, even if it is difficult to obtain the explicit form of the bounce solution,
our bound is useful as a quick check on the stability of the false vacuum.
In section \ref{sec:lower}, we discuss the lower bound on the minimal bounce action.
In section \ref{sec:comparisons}, we compare our lower bound
with the actual value or the upper bound for some representative examples.

\section{An lower bound on the bounce action}\label{sec:lower}
Here, we derive an absolute lower bound on the bounce action. 
We consider $m$ scalar fields with the canonical kinetic term in $N$-dimensional Euclidean space.
The action is given by
\begin{align}
         S[\vec{\phi}]&=T[\vec{\phi}]-U[\vec{\phi}],\\
         T&=
         \int d^Nx~ \sum_{a=1}^m\sum_{i=0}^{N-1}\frac{1}{2}\left(\frac{\partial\phi_a}{\partial x_i}\right)^2,\\
         U&=\int d^Nx~U(\phi),
\end{align}
where $\vec{\phi}\equiv(\phi_1,\phi_2,..,\phi_m)$ and $U$ is the inverted potential: 
$U(\Phi)\equiv-V(\Phi)$
with $V(\Phi)$ being the actual one.
Throughout this Letter,
we set the false vacuum at $\vec{\phi}=0$ (and $V(\vec{0})=0$) without loss of generality.

Let us consider $N=4$ dimensional Euclidean space for a while.
If $\phi$ is a solution of equation of motion, it stationalizes the action.
Considering the rescaling of the Euclidean space coordinates $\phi(x)\rightarrow \phi(\xi x)$,
we have the following relation
\begin{align}
         \left.\frac{\partial S[\phi(\xi x)]}{\partial \xi}\right|_{\xi=1}=
        -2T+4U=0,
\end{align}
which leads to
\begin{align}
         S=\frac{T}{2}.
\end{align}
Thus, the problem of finding the minimal action solution can be reduced to
that of finding the minimal kinetic energy solution.
Since the minimal action bounce solution is known to be
O$(N)$ symmetric for $N>2$ and
even for multiscalar cases~\cite{Coleman:1977th, lopes1996radial, byeon2009symmetry, Blum:2016ipp},
we consider an O$(4)$ symmetric bounce whose radial coordinate is $r$.
With O$(4)$ symmetry, the kinetic energy $T$ is given by
\begin{align}
         T\equiv \sum_{a=1}^m \int_0^{\infty} dr~\pi^2 r^3\dot{\phi}_a^2.
\end{align}
The equation of motion is 
\begin{align}
         \ddot{\phi}_a + \frac{3}{r}\dot{\phi}_a + \frac{\partial U}{\partial \phi_a} = 0 \quad(a=1,\cdots,m), \label{eq:eom}
\end{align}
where we denote the ``dot'' as a derivative with respect to $r$. 

To discuss the minimum kinetic energy $T$, we define a class of bounce solutions.
We characterize them by two parameters:
field difference $\Delta \phi_a \equiv \phi_a(0) - \phi_a(\infty)$
and potential difference $\Delta U \equiv U[\phi(0)] - U[\phi(\infty)]$.
Our first goal is to derive a lower bound for such a class of solution.
We can easily see $\Delta\phi_a$ and $\Delta U$ are functional of $\dot\phi_a$'s.
By multiplying $\dot{\phi}_a$ to Eq.~(\ref{eq:eom}) and integrating from zero to infinity, we obtain
\begin{align}
         \left[\sum_{a=1}^m \dot{\phi}_a^2/2 + U\right]^{r=0}_{r=\infty} = \Delta U = \sum_{a=1}^m \int_0^\infty dr \frac{3\dot{\phi}_a^2}{r}.
\end{align}
On the other hand,
\begin{align}
         \Delta \phi_a = -\int_0^\infty dr \dot{\phi}_a,
\end{align}
holds.
To consider the minimization problem on $T$ with fixed $\Delta \phi_a$ and $\Delta U$,
we introduce the Lagrange multiplier $\alpha_a$ and $\beta$,
and define $\tilde T$ as
\begin{align}
\tilde T[\phi, \{\alpha_a\}, \beta]
= & T[\phi] + \sum_{a=1}^m 2\alpha_a \left( \Delta \phi_a + \int_0^\infty dr \dot\phi_a \right) \nonumber\\
  & \quad - \beta \left( \Delta U - \sum_{a=1}^m \int_0^\infty dr \frac{3\dot\phi_a^2}{r} \right).
\end{align}
An extremum condition $\delta\tilde T / \delta\dot\phi_a = 0$ gives
\begin{align}
\dot\phi_a = -\frac{\alpha_a r}{\pi^2 r^4 + 3\beta} \qquad (a=1,\cdots,m). \label{eq:phisol}
\end{align}
In the above solution, the Lagrange multiplier $\alpha_a$ and $\beta$ are determined from
the constraints $\int_0^\infty dr \dot\phi_a = -\Delta\phi_a$ and $\sum_{a=1}^m \int_0^\infty dr (3/r)\dot\phi_a^2 = \Delta U$ as
\begin{align}
\alpha_a = \frac{24\Delta\phi_a|\Delta\phi|^2}{\Delta U}, \qquad
\beta = \frac{12|\Delta\phi|^4}{\Delta U^2}, \label{eq:absol}
\end{align}
where $|\Delta\phi| = \sqrt{\sum_{a=1}^m \Delta\phi_a^2 }$.
At this point, the solution Eqs.~(\ref{eq:phisol}, \ref{eq:absol}) is just an extremum,
and it is not clear whether this point is the global minimum or not.
To check this point, let us see $\tilde T$ again with Eq.~(\ref{eq:absol}).
\begin{align}
& \tilde T\left[\phi,~ \left\{\alpha_a=\frac{24\Delta\phi_a|\Delta\phi|^2}{\Delta U}\right\},~ \beta=\frac{12|\Delta\phi|^4}{\Delta U^2} \right] \nonumber\\
=& \frac{12|\Delta\phi|^4}{\Delta U}
   + \sum_{a=1}^m \int_0^\infty \left( \frac{\pi^2 r^4 + 3\beta}{r} \right)  \left( \dot\phi_a + \frac{\alpha_ar}{\pi^2 r^4 + 3\beta} \right)^2.
\end{align}
The above equations tells us that the solution Eqs.~(\ref{eq:phisol}, \ref{eq:absol}) does give the global minimum on $T$
for fixed $\Delta\phi_a$ and $\Delta U$.

Then, we can write the following inequality on the bounce action $S$ (if it exists)
by using $|\Delta\phi|$ and $\Delta U$ as
\begin{align}
S \geq \frac{24}{\lambda_\phi(\Delta\phi_a)},\qquad
\lambda_\phi(\Delta\phi_a) \equiv \frac{4\Delta U}{|\Delta\phi|^4}. \label{eq:bound1}
\end{align}
The above inequality is saturated if and only if $\dot\phi_a = -\alpha_a r / (\pi^2 r^4 + 3\beta)$ holds\footnote{
One may be interested in the potential which realizes the bounce solution $\dot\phi = -\alpha_a r / (\pi^2 r^4 + 3\beta)$.
The explicit form of the potential for a single scalar field case is
$U = \Delta U[ \phi/\Delta \phi - (4/3\pi)\sin(\pi\phi/\Delta\phi) + (1/6\pi)\sin(2\pi\phi/\Delta\phi)]$.
This potential give the minimum of the bound Eq.~(\ref{eq:bound1}).
However, we do not know an example which saturates the bound Eq.~(\ref{eq:bound2}).
Thus, the bound Eq.~(\ref{eq:bound2}) may be weaker than Eq.~(\ref{eq:bound1}).
}.
To calculate this bound, we need $\Delta\phi_a$, \textit{i.e.}, $\phi_a(r=0)$.
Although we do not know about $\Delta\phi_a$ unless we explicitly solve the equation of motion,
we can set a bound on the minimal action even without solving the equation of motion.
Suppose there exists $\lambda (>0)$ such that
\begin{align}
         -U(\phi_a) = V(\phi_a)\geq -\frac{\lambda}{4}|\phi|^4. \label{eq:def_of_lambda}
\end{align}
We can find $\lambda$ for the potential such that $V(\phi)/|\phi|^4$ is bounded below.
Then, we can define $\lambda_{\rm cr}$ as
\begin{align}
\lambda_{\rm cr} \equiv {\rm max}\left[ \frac{-4V(\phi^a)}{|\phi|^4}\right]. \label{eq:def_of_lambdacr}
\end{align}
We can see this $\lambda_{\rm cr}$ is the minimum of a set of $\lambda$ which satisfy Eq.~(\ref{eq:def_of_lambda}).
Then, the bounce action has an absolute lower bound
\begin{align}
S \geq \frac{24}{\lambda_{\rm cr}}. \label{eq:bound2}
\end{align}
because $\lambda_\phi \leq \lambda_{\rm cr}$ is satisfied for any value of $\phi_a$.
As a reference, the Fubini instanton~\cite{Fubini:1976jm},
which is a bounce solution with negative quartic potential $V = (\lambda_4/4) \phi^4$,
has $S = 8\pi^2/3\lambda_4 \simeq 26.3/\lambda_4 >24/\lambda_{\rm cr}$
because $\lambda_{\rm cr} = \lambda_4$ holds in this case. 
To derive the above bound Eq.~(\ref{eq:bound2}), we do not need the explicit form of the bounce solution.
Although the bound Eq.~(\ref{eq:bound2}) may be weaker than Eq.~(\ref{eq:bound1}),
we can derive Eq.~(\ref{eq:bound2}) \textit{only} from the information of the potential.
In $N(>2)$ dimensional case, the same procedure gives the lower bound:
\begin{align}
         S \geq \frac{4\left[N(N-1)(N-2)\right]^{\frac{N-2}{2}}}{N\Gamma(N/2)}
         \sin(2\pi/N)^{\frac{N}{2}}\left(\frac{1}{\lambda_N}\right)^{\frac{N-2}{2}},
\end{align}
with $\lambda_N \equiv N\Delta{U}/(\Delta\phi)^{\frac{2N}{N-2}}$.

One may be interested in the condition in which the lower bound Eq.~(\ref{eq:bound2}) becomes close to the actual value.
As long as the true vacuum and the false vacuum are not degenerated, our method can give a good estimation on the lower bound of the decay rate.
For detailed discussion, see the Appendix.

So far, we have derived a lower bound on the bounce action.
Here let us comment on an upper bound on the bounce action.
As is discussed above,
by finding a point $\phi_{\rm cr}$ which maximizes $ [ -4V(\phi) /|\phi|^4] $,
we can obtain a lower bound on the action.
By using this $\phi_{\rm cr}$,
we can easily obtain an upper bound as discussed in Ref.~\cite{Dasgupta:1996qu}.
First, we restrict the field space into $\phi_{\rm cr}$ direction, which
is a straight line passing through the false vacuum $\phi=0$ and $\phi_{\rm cr}$.
We obtain a reduced single field theory on this straight line,
and we can easily estimate the bounce action of this reduced action.
Then, the resulting bounce action becomes an upper bound
on the actual minimal bounce action.
Thus, by
 finding a point $\phi_{\rm cr}$ which maximizes $ [ -4V(\phi) /|\phi|^4] $ ,
we can obtain both a lower and an upper bound on the actual minimal bounce action at the same time.

Now, let us briefly discuss the applicability of our results.
As long as the kinetic term is canonical
and once the potential of the scalar fields is determined,
our method gives a lower bound on the classical bounce action in a simple way.
In some models, quantum corrections or thermal loop corrections are essential to generate a barrier between the false and the true vacuum.
In such cases, the effective potential can be used for our method,
and our method gives a good estimation on the lower bound of the bounce action as long as the perturbative calculation around the bounce can be used.
In general,
to obtain the vacuum decay rate precisely,
we need to estimate the prefactor by integrating out fluctuations around the actual bounce
solution\footnote{
The gauge dependence and renormalization scale dependence
are canceled by considering loop corrections \cite{Endo:2017gal, Endo:2017tsz}.}:
\begin{align}
         \frac{\Gamma}{V}=A'\mu^{4} \exp[{-S_{\rm cl}}]=\mu^{4}\exp[{-S_{\rm cl}+\ln A'}],
\end{align}
 where  $\Gamma/V$ denotes decay rate per unit four-dimensional volume,
 $\mu$ denotes a typical energy scale of the bounce dynamics, 
 $S_{\rm cl}$ denotes classical bounce action,
 and $A'$ is a (normalized) prefactor.
As long as the theory is perturbative, we may expect $\ln A' \sim\mathcal{O}(1)$ although there is a little ambiguity on the definition of $\mu$.
 In the case of $S_{\rm cl}\gg {\cal O}(1)$,
 the vacuum decay rate is mainly determined by the classical bounce action
 and our bound becomes useful to determine the order of the decay rate.
Actually, when we consider the cosmological history of the
vacuum, the relevant range of the action is $S_{\rm cl}\sim\mathcal{O}(100)$.
Also, the condition for the thermal transition in the expanding Universe is $H^4 \sim T^4 e^{-S_3/T}$,
where $H$ is the Hubble expansion rate and $T$ is temperature.
The typical size of dimensionless action $S_3/T$ is ${\cal O}(100)$.
In these cases, our bound can provide a lower bound on the vacuum decay rate.

Finally, we derive a sufficient condition of vacuum stability in our present Universe.
In order to have a stable Universe,
the vacuum decay should not happen within a Hubble volume in a Hubble time:
\begin{align}
\label{eq:cond:vac}
         H_0^4 \gtrsim \Gamma/V,
\end{align}
where $H_0\sim 10^{-42}~$GeV is the Hubble constant today.
On the other hand, by using Eq.~(\ref{eq:bound1}), the vacuum decay rate per volume to the point ${\phi}_a$ (if it exists) is bounded as
\begin{align}
\label{eq:cond:min}
         \frac{\Gamma({\phi}_a)}{V} \lesssim |\phi|^4 \exp\left( - \frac{6|\phi|^4}{-V(\phi_a)} \right),
\end{align}
where we assume that the size of prefactor is roughly given by $|\phi|^4$.
By using Eqs.~(\ref{eq:cond:vac}) and (\ref{eq:cond:min}),
we can show a sufficient condition of vacuum stability
on the shape of the potential:
\begin{align}
         V(\phi) + \frac{1}{4} \lambda_{H_0}(|\phi|) |\phi|^4>0, \label{eq:potential_bound}
       \end{align}
with\footnote{
Strictly speaking,
this condition should be imposed in $\phi>H_0$ region.
}
\begin{align}
           \lambda_{H_0}(|\phi|)&=\frac{3}{\ln(|\phi|^2/H_0^2)}\nonumber\\
           &\simeq
         \frac{3}{\ln(|\phi|^2/1\text{ GeV}^2)+193}. 
\end{align}
This is a sufficient condition for the stability of a false vacuum.
Any false vacuum with any potential which satisfies Eq.~(\ref{eq:potential_bound})
has a lifetime which is longer than the age of the Universe.

\section{Comparisons with the actual value}\label{sec:comparisons}
In this section, we discuss several explicit examples in four-dimensional space.
We will see the consistency of Eq.~(\ref{eq:bound2}),
and furthermore, see that the lower bound Eq.~(\ref{eq:bound2}) becomes close to the actual value of the bounce action in many cases.
In this sense, the lower bound Eq.~(\ref{eq:bound2}) is a quite useful tool to estimate the value of the bounce action
when an explicit calculation is difficult.

\subsection{Single scalar field}
\begin{figure}[h!]
\centering
\includegraphics[width=0.9\hsize]{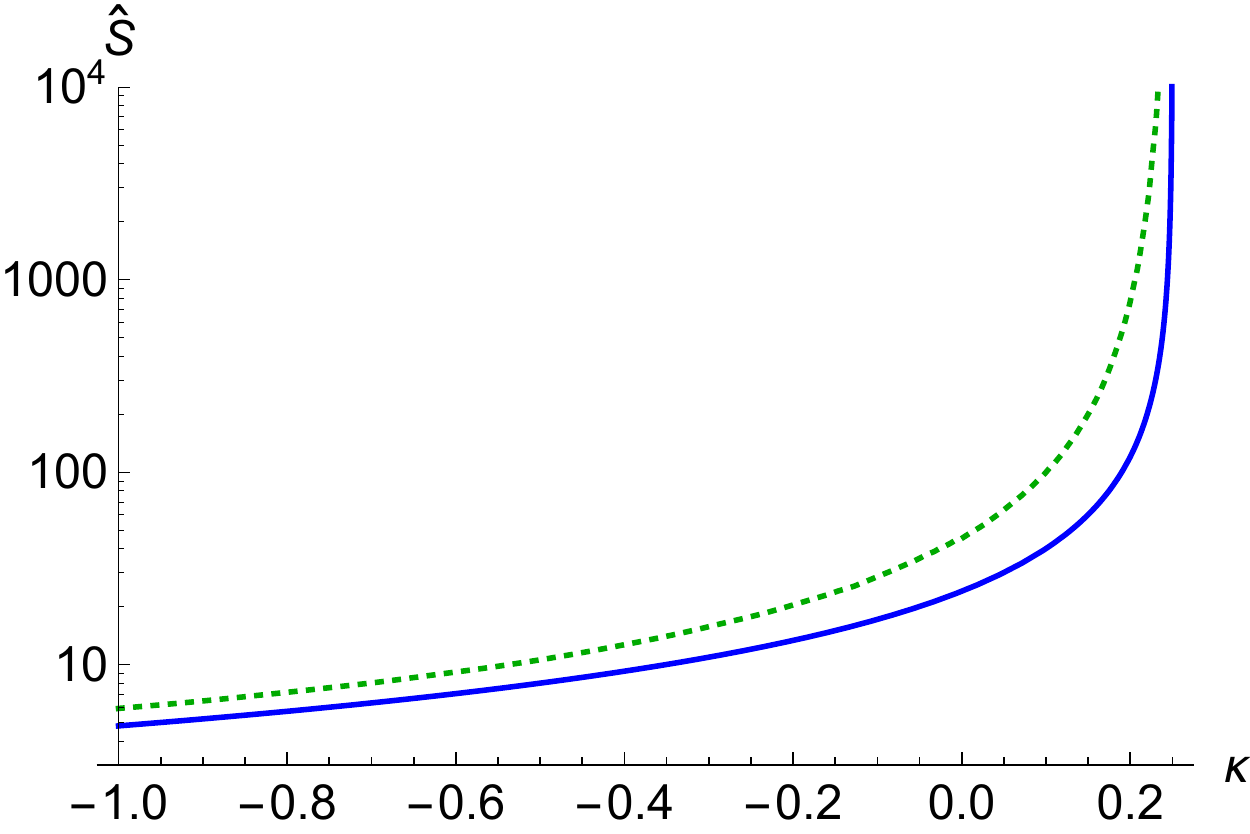}
\caption{
The minimal bounce action and the lower bound.
The blue solid line shows the lower bound which is given in Eq.~(\ref{eq:bound2}).
The green dotted line is taken from Fig.~1 in Ref.~\cite{Sarid:1998sn}.
}
\label{fig:comparison with sarid}
~\\[5mm]
\includegraphics[width=0.9\hsize]{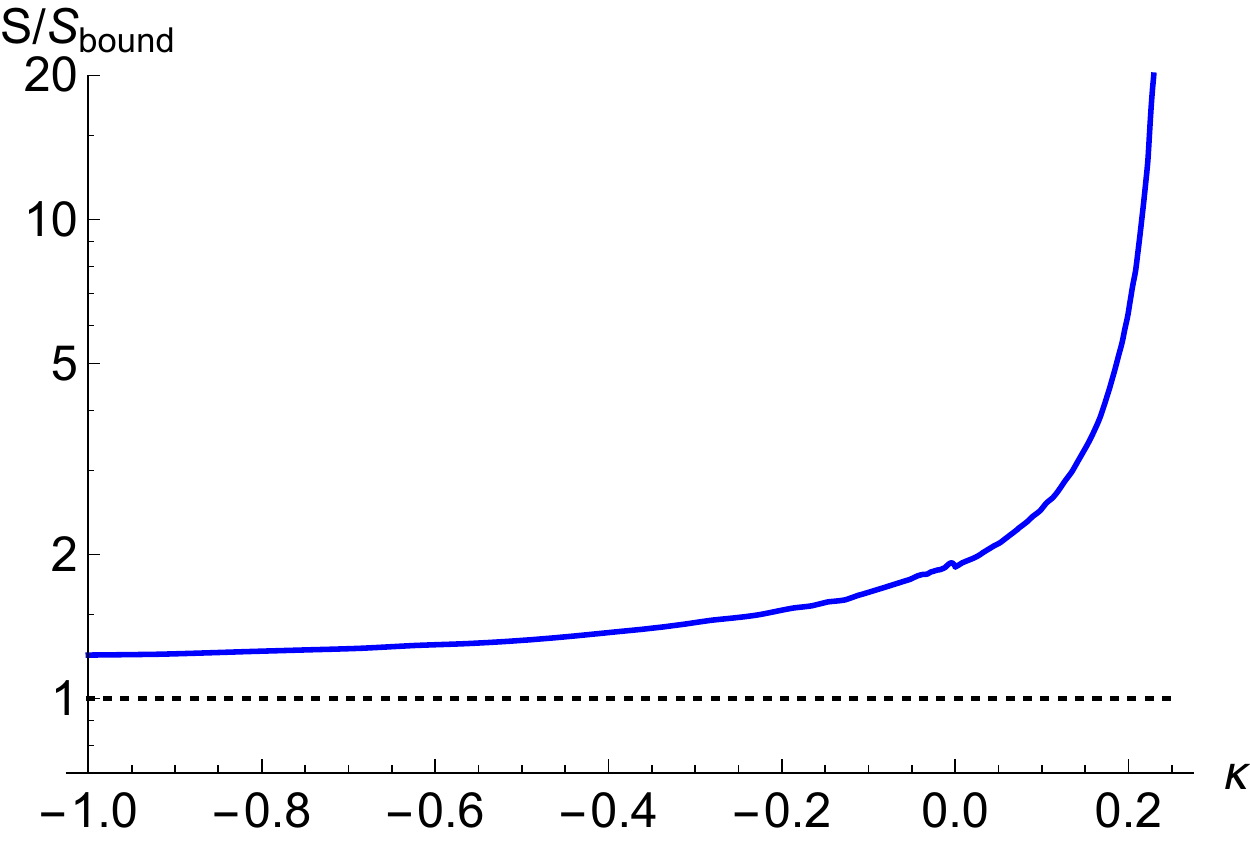}
\caption{
The ratio between the green dotted line and the blue solid line in Fig.~\ref{fig:comparison with sarid}.
}\label{fig:comparison with sarid (ratio)}
\end{figure}
The first example is a single scalar field theory with a polynomial potential:
\begin{align}
V(\phi) = \frac{1}{2}M^2 \phi^2 - \frac{1}{3} A \phi^3 + \frac{1}{4} \lambda_4 \phi^4.
\end{align}
This potential gives us good insight into a relationship
between our bound Eq.~(\ref{eq:bound2}) and the minimal bounce action $S$.
As discussed in Ref.~\cite{Sarid:1998sn}, we can parametrize the minimal bounce action as
\begin{align}
S = \frac{9M^2}{2A^2} \hat{S}(\kappa),\qquad
\kappa \equiv \frac{9 \lambda_4 M^2}{8 A^2}.
\end{align}
Here $\hat S$ is a function which only depends on $\kappa$.
According to the definition given in Eq.~(\ref{eq:def_of_lambdacr}), $\lambda_{\rm cr}$ is calculated as
\begin{align}
\label{eq:lamcrsingle}
         \lambda_{\rm cr}=\frac{2A^2}{9M^2}-\lambda_4.
\end{align}
By using this $\lambda_{\rm cr}$, we obtain the bound on $\hat S(\kappa)$ as
\begin{align}
\hat{S}(\kappa) \geq \frac{24}{1-4\kappa}. \label{eq:bound on shat}
\end{align}
Ref.~\cite{Sarid:1998sn} gives the numerical result of $\hat S(\kappa)$ by calculating the bounce configuration,
and we show a comparison between the result of Ref.~\cite{Sarid:1998sn}
and the bound Eq.~(\ref{eq:bound on shat}) in Fig.~\ref{fig:comparison with sarid} and Fig.~\ref{fig:comparison with sarid (ratio)}.
We can see that our bound becomes close for large negative $\kappa$.
In this regime, the bounce solution is well described by the Fubini instanton \cite{Fubini:1976jm}.
On the other hand, our bound departs from the numerical value of the minimal bounce action if $\kappa$ is close to $1/4$,
in this regime, the false and true vacua are almost degenerate and the bounce solution is well described by thin-wall approximation.
There exists a potential barrier between the true vacuum and false vacuum.

\subsection{Multi scalar fields}
\begin{table}
\centering
\begin{tabular}{|r||c|c|c|c|}
\hline
                & $m=1$ & $m=2$ & $m=4$ & $m=8$ \\\hline\hline
$1  \leq R<1.2$ & 0.16 & 0.52 & 0.78 & 0.96 \\\hline
$1.2\leq R<1.5$ & 0.22 & 0.26 & 0.13 & 0.02 \\\hline
$1.5\leq R<2$   & 0.14 & 0.12 & 0.09 & 0.01 \\\hline
$2  \leq R<5$   & 0.09 & 0.04 & 0.01    & 0 \\\hline
$5  \leq R$     & 0.02 & 0.01 & 0    & 0 \\\hline
Stable          & 0.37 & 0.04 & 0    & 0 \\\hline
\end{tabular}
\caption{
The distribution of $R\equiv S_{\rm upper}/S_{\rm lower}$ for the multiscalar potential.
}\label{tab:multiscalar}
\end{table}
The second example is a polynomial potential with multiscalar field $\phi_1,...,\phi_m$.
We consider a term up to the quartic interaction,
and parametrize it as follows
\begin{align}
         V=\sum_{i}M^2\mu_i \phi_i^2+\sum _{i,j,k}M\gamma_{ijk}\phi_i\phi_j\phi_k
         +\sum_{ijkl}\lambda_{ijkl}\phi_i\phi_j\phi_k\phi_l,
\end{align}
where $M$ is a mass scale which does not affect the value of classical action
and
$\mu_i,\gamma_{ijk},\lambda_{ijkl}$ denote some dimensionless coupling.
Here we do not calculate the bounce configuration explicitly.
Instead of the explicit calculation, we estimate a lower and upper bound on the bounce action.
The upper bound is estimated by the straight line method described in the later part of Sec.~\ref{sec:lower}.
We define the ratio between the upper and lower bound as $R\equiv S_{\rm upper}/S_{\rm lower}$.
If this $R$ is close to $1$, our lower bound is close to the actual value of the bounce action.

We calculate the ratio $R$ by taking $\mu$'s, $\gamma$'s, and $\lambda$'s as random variables as in Ref.~\cite{Dine:2015ioa}.
The ranges of the parameters are taken as
\begin{align}
         0<\mu_i<1,~~
         -\frac{1}{m}<\gamma_{ijk}<\frac{1}{m},~~
         -\frac{1}{m}<\lambda_{ijkl}<\frac{1}{m}.
\end{align}
We take the range of $\gamma$ and $\lambda$ so that the theory remains stable against loop corrections. 
We generate 1000 parameter points, and show the distribution of $R$ in Tab.~\ref{tab:multiscalar}.
This result shows the lower bound Eq.~(\ref{eq:bound2}) becomes close to the actual value of the bounce action in the case of a large number of scalar fields.
This feature can be understood as follows.
As we have seen in the previous single scalar example,
$\lambda_{\rm cr}$ depends on quartic coupling and the cubic coupling square
 (see Eq.~(\ref{eq:lamcrsingle})).
In the present case, the typical value of 
quartic coupling is $1/m$ and that of cubic coupling square is $1/m^2$.
With larger $m$, quartic coupling becomes more and more relevant
and the bounce action becomes close to our lower bound.

\subsection{MSSM}
\begin{figure}
\centering\includegraphics[width=0.9\hsize]{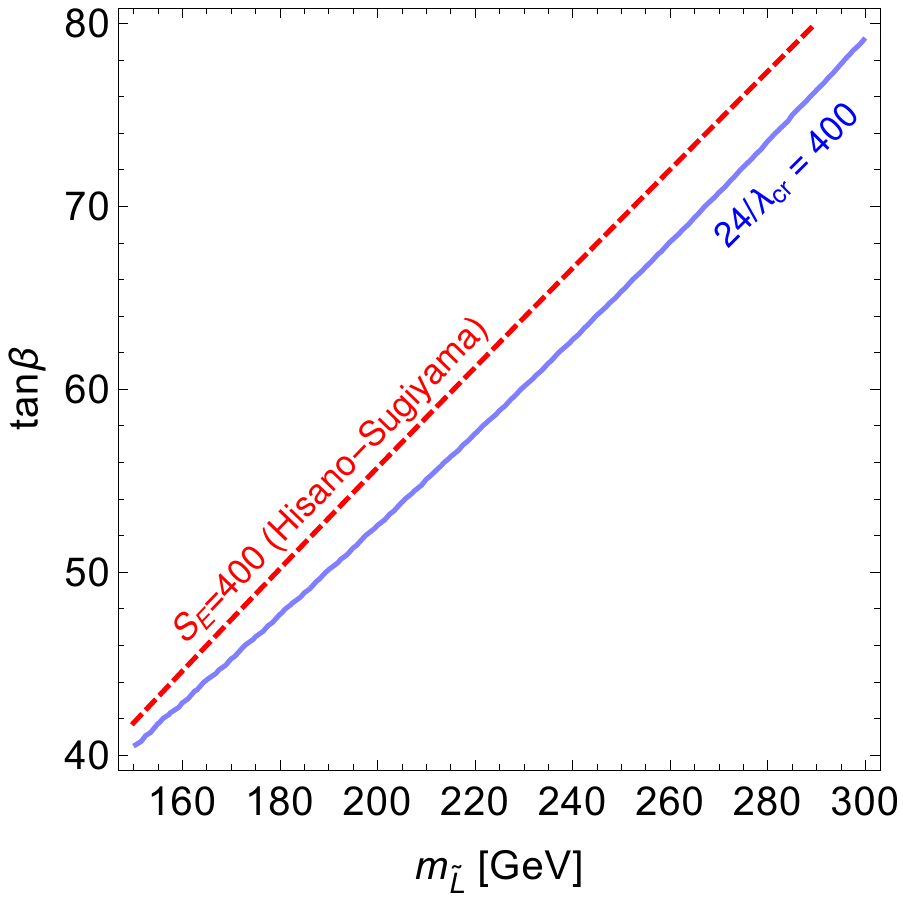}
\caption{
Vacuum stability constraints on the $m_{\tilde L}$-$\tan\beta$ plane.
Here we take $\mu = 700~{\rm GeV}$ and $m_{\tilde\tau_R} = m_{\tilde L} + 200~{\rm GeV}$.
The blue line is written by using the RHS of Eq.~(\ref{eq:bound2}).
The red dashed line is written by using a fitting formula given in Ref.~\cite{Hisano:2010re}.
}\label{fig:stau}
\end{figure}
The last example is the MSSM.
Supersymmetric models introduce a lot of scalar partners of the standard model fermions,
and sometimes they destabilize the standard model-like vacuum.
For example, Ref.~\cite{Hisano:2010re} discussed a vacuum stability in a direction of the third generation slepton with large $\tan\beta$.
The scalar potential for the up-type Higgs $H_u$, the left-handed stau $\tilde L$, and the right-handed stau $\tilde\tau_R$ is given as
\begin{align} 
V =
& (m_{H_u}^2 + \mu^2) |H_u|^2
+ m_{\tilde L}^2 |\tilde L|^2
+ m_{\tilde \tau_R}^2 |\tilde \tau_R|^2 \nonumber\\
& + \frac{g_2^2}{8}( |\tilde L|^2 + |H_u|^2 )^2
+ \frac{g_Y^2}{8}( |\tilde L|^2 - 2|\tilde\tau_R|^2 - |H_u|^2 )^2 \nonumber\\
& + \frac{g_2^2 + g_Y^2}{8} \delta_H |H_u|^4
+ y_\tau^2 |\tilde L\tilde \tau_R|^2 \nonumber\\
& - (y_\tau \mu H_u^* \tilde L \tilde\tau_R + h.c.).
\end{align}
Here we do not consider the down-type Higgs $H_d$ because its VEV is suppressed by $1/\tan\beta$.
$\delta_H$ expresses a radiative correction from the top quark and the stop, and its typical value is $\delta_H \simeq 1$.
A cubic term $H_u^* \tilde L \tilde\tau_R$ in the last line destabilizes the standard model-like vacuum.
Its coupling constant is proportional to $\mu\tan\beta$.
In Fig.~\ref{fig:stau}, we show a comparison between the lower bound on the bounce action which is given in Eq.~(\ref{eq:bound2})
and Ref.~\cite{Hisano:2010re}.
The lower bound on the bounce action $S$ is 400 at the blue line, and the standard model-like vacuum is sufficiently stable in the lower right region of the blue line.
By using the result in Ref.~\cite{Hisano:2010re}, in Fig.~\ref{fig:stau}, we show the red line on which $S=400$ is satisfied.
We can see our bound Eq.~(\ref{eq:bound2}) is consistent with the result of Ref.~\cite{Hisano:2010re}.

To discuss the stability in the upper left region, Eq.~(\ref{eq:bound2}) is not enough in general.
However, Fig.~\ref{fig:stau} shows that
the sufficient stability condition by the blue line only differs by 5 \% from the upper bound on $\tan\beta$ by the red line.
This means that Eq.~(\ref{eq:bound2}) gives a good estimation on the upper bound of $\tan\beta$.
Actually, Figs.~\ref{fig:comparison with sarid}, \ref{fig:comparison with sarid (ratio)} show
the lower bound on the bounce action gives a good estimation on the actual value unless the true and false vacua are degenerated.
Such a degenerated situation is a special situation in the sense that it requires a tuning of the parameters or an approximate symmetry between two vacua.
Thus, we can expect that our discussion is useful to discuss more complicated models.

\section{Conclusion}\label{sec:conclusion}
In this Letter, we derived a generic lower bound Eq.~(\ref{eq:bound2}) on the bounce action
by using a quite simple discussion with the Lagrange multiplier.
Our bound can be applied to a broad class of scalar potential with any number of scalar field.
Necessary information to derive this bound is only $\lambda_{\rm cr}$ which is defined by Eq.~(\ref{eq:def_of_lambdacr}).
In particular, our bound provides useful information for a model with a large number of scalar fields such as the landscape scenario
because we do not need the explicit form of the bounce solution.
By using this result, in Eq.~(\ref{eq:potential_bound}),
we derived a sufficient condition of the stable vacuum of the Universe for a general scalar potential.
The bound Eq.~(\ref{eq:potential_bound}) can be used as a quick check on the stability of a false vacuum 
in a broad class of models.
As we discussed in section \ref{sec:comparisons}, the lower bound Eq.~(\ref{eq:bound2}) gives a good estimation on the actual value in many cases.
We also investigated a condition for when the bounce action becomes close to the lower bound.
As long as two vacua are not almost degenerated 
the minimal bounce action can be close to the lower bound.
We have seen this feature in some representative examples. 

\section*{Acknowledgements}
We thank Kfir Blum for fruitful discussions and careful reading of the Letter.
We are also grateful to Sonia Paban for useful comments.
We also thank Junji Hisano for a clarification of Ref.~\cite{Hisano:2010re}.
The work of MT is supported by the JSPS Research Fellowship for Young Scientists.


\appendix
\section{Criteria for kinetic bounce solution}\label{sec:criteria}
Here, we discuss the situation in which our bounce becomes close to the actual value of the minimal bounce action.
First, let us look at large $r$ behaviour of the field configuration which gives the lower bound
(see Eq.~(11) in the main text).
It is given by
\begin{align}
         \phi\propto \frac{1}{r^2},~~(\text{for }r^2\gg\Delta\phi^2/\Delta U). 
\end{align}
Note that the bounce solution behaves as $\phi \sim e^{-mr}$ for large $r$ if there exists a mass term around the false vacuum,
and the above solution is a solution of equation of motion without potential term:
\begin{align}
         \ddot{\phi}+\frac{3}{r}\dot{\phi}=0.
\end{align}
We denote this class of bounce solutions as \textit{``kinetic bounce solution''}.
Thus, it seems that if the potential term becomes ineffective in large $r$,
the bounce action can take a value close to the lower bound.
Below, we will show this is actually the case.
In addition, we derive a necessary condition for that the bounce dynamics at large $r$ is dominated by the kinetic term. 
This condition is necessary in order for the bounce action to be close to the lower bound.

To see a condition for that the kinetic term dominates,
it is instructive to consider the following simple (linear+flat) potential:
\begin{align}
         V(\phi)=
         \begin{cases}
         0 & (\phi<\phi_*),\\
         (\phi_*-\phi) F & (\phi>\phi_*),
         \end{cases} \label{eq:linearflat}
\end{align}
where $\phi_* (>0)$ and $F(>0)$ are some constants.
The minimal $\lambda$, which satisfies $V+\lambda \phi^4/4\geq0$, is given by
\begin{align}
         \lambda_{\rm cr}=\left(\frac{3}{4\phi_*}\right)^3F. \label{eq:lambdamin}
\end{align}
We define ``false vacuum'' at $\phi=0$.
The bounce solution is uniquely determined and given by
\begin{align}
         \phi(r)=\begin{cases}
         \frac{r_*^2-r^2}{8}F+\phi_*& (r<r_*),\\
         \phi_*\left(\frac{r_*}{r}\right)^2&(r>r_*),
         \end{cases} \label{eq:kinetic_bounce}
\end{align}
with
\begin{align}
        r_*^2=8\frac{\phi_*}{F}. \label{eq:rstar}
\end{align}
Note that for $\phi<\phi_*$, the kinetic term fully determines the dynamics.
To distinguish the bounce action which will be discussed later,
we denote the bounce action under the potential Eq.~(\ref{eq:linearflat}) as $S_0$.
The bounce action $S_0$ is slightly larger than the lower bound $24/\lambda_{\rm cr}$ and given by
\begin{align}
\label{eq:exsam}
       S_0 = \frac{45\pi^2}{16\lambda_{\rm cr}}\simeq \frac{27.7}{\lambda_{\rm cr}}.
\end{align}

Next, let us add a potential term $V_+$ in $0\leq\phi<\phi_*$ region,
and see how $V_+$ changes the value of the minimal action $S$.
If the potential energy satisfies $|V_+| \ll \dot\phi^2$,
the bounce solution obeys $\ddot\phi + (3/r)\dot\phi = 0$.
In this case, the kinetic energy at $r>r_*$ is written as
\begin{align}
         \dot{\phi}^2(r)\sim \frac{1}{4}\lambda_{\rm cr}\phi^3(r) \phi_*.
\end{align}
The condition $|V_+| \ll \dot\phi^2$ is broken at $\phi = \phi_c$ such that
\begin{align}
       \frac{1}{4} \lambda_{\rm cr}\phi_c^3\phi_* \sim V_+(\phi_c). \label{eq:phic}
\end{align}
We define radius $r_c$ such that $\phi(r_c) = \phi_c$.
By definition, $r_c$ is larger than $r_*$.
We split the bounce action $S$ as
\begin{align}
S = S_{r>r_c} + S_{r<r_c},
\end{align}
where $S_{r>r_c} \equiv \int_{r_c}^\infty dr \pi^2 r^3 [ \dot\phi^2 + 2U(\phi)]$ and $S_{r<r_c} \equiv S - S_{r>r_c}$.
For $r>r_c$, we cannot neglect the potential term in the equation of motion.

First, let us see an effect on $S_{r<r_c}$. 
By using Eq.~(\ref{eq:kinetic_bounce}) and Eq.~(\ref{eq:phic}),
we can estimate the ratio of $r_*$ over $r_c$ and $\phi_*$ over $\phi_c$ as
\begin{align}
         \frac{r_*}{r_c}&\sim \left(\frac{V_+(\phi_c)}{\Delta U}\right)^{1/6},\qquad
         \frac{\phi_*}{\phi_c}&\sim \left(\frac{V_+(\phi_c)}{\Delta U}\right)^{-1/3}. \label{eq:ratio_star_c}
\end{align}
As long as $V_+(\phi_c)\ll{\Delta U}$, by a tiny shift of initial position 
$\phi(0)\rightarrow \phi(0)+\delta$ with $\delta\sim\phi_c\ll\phi(0)$,
the kinetic energy at $\phi_c$ changes by a factor and
we will obtain a bounce solution.
In this case, $S_{r<r_c}$ remains almost the same:
\begin{align}
        \frac{S_{r<r_c}}{S_0} - 1
        ~\sim~ \mathcal{O}\left(\left(\frac{V_+(\phi_c)}{\Delta U}\right)^{1/3}\right).
\end{align}

Next, let us see an effect on $S_{r>r_c}$.
We denote the maximal value of the potential inside $\phi_c$
to be $V_{\rm max}(\phi_c) = {\rm max}\{V(\phi) | 0\leq \phi \leq \phi_c\}$.
A typical mass scale at $0\leq \phi \leq \phi_c$ may be
given by $m_{\rm typ}^2\sim V_{\rm max}/\phi_c^2$.
Then, the bounce action inside $\phi_c$ would be estimated as 
\begin{align}
         S_{r>r_c} \sim T_{r>r_c}
         &\sim \int_{r_c}^{r_c+m^{-1}_{\rm typ}}dr~r^3\dot{\phi}^2\nonumber \\
         &\sim m_{\rm typ}\phi_c^2r_c^3 \nonumber \\
         &\sim \frac{m_{\rm typ}r_*^2}{r_c} S_0.
\end{align}
We can see as long as
\begin{align}
\label{eq:condmax}
         \frac{m_{\rm typ}r_*^2}{r_c}\ll1,
\end{align}
the contribution from $S_{r>r_c}$ is suppressed.
The left hand side of Eq.~(\ref{eq:condmax}) is small if we consider small and mild shape of $V_+$.
We conclude a necessary condition for the minimal action which is close to the lower bound is
\begin{align}
V_{\rm max} \ll \Delta U.
\end{align}
If this condition is violated, the minimal action $S$ is significantly deviated from $S_0$.
\begin{figure}[h!]
\centering
\includegraphics[width=0.74\hsize]{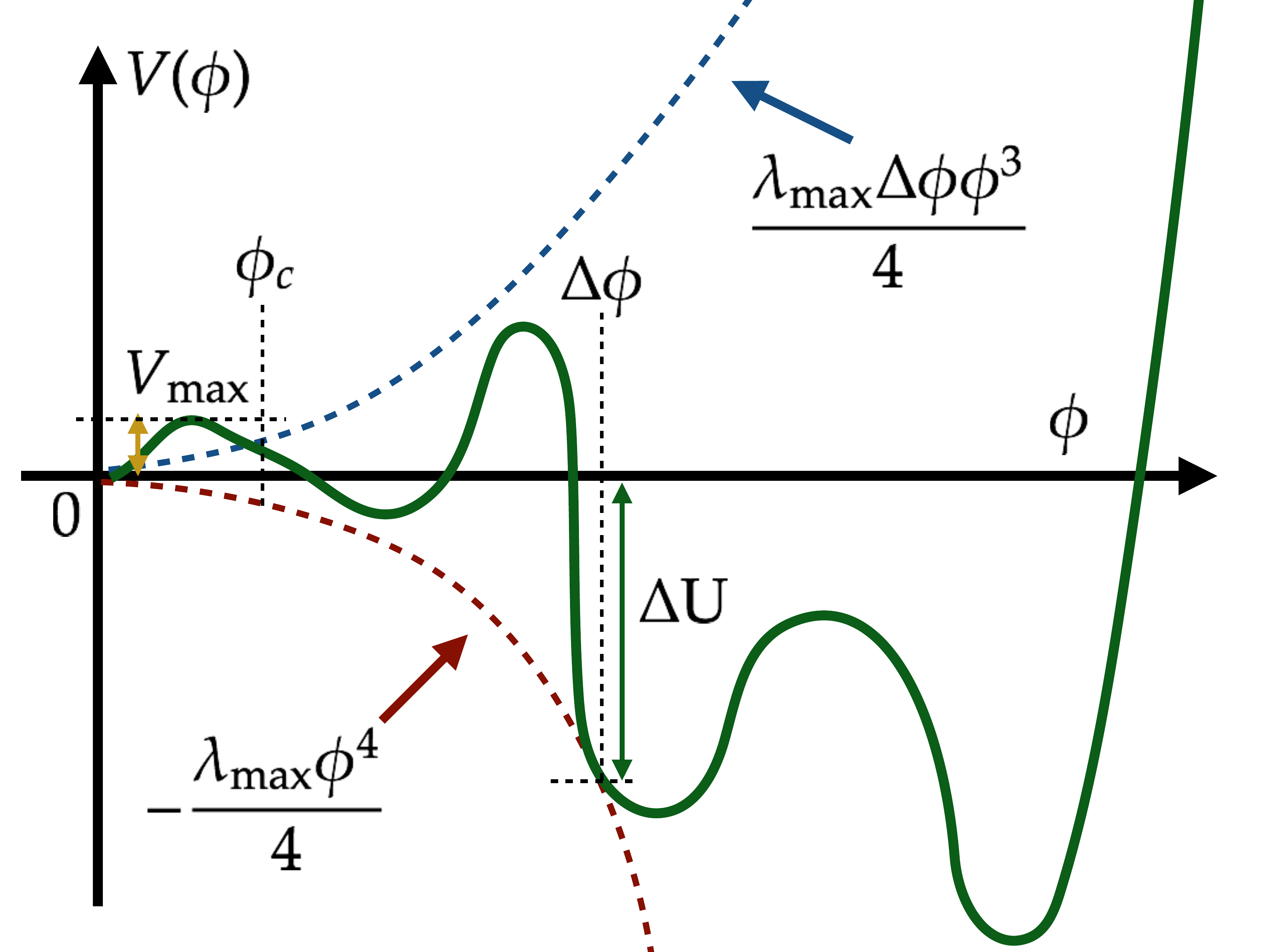}
\caption{
A schematic picture of our procedure.
}
\label{fig}
\end{figure}
Now, let us generalize the previous discussion.
For given potential $V(\phi)$, we can define $\lambda_{\rm cr}$ as a minimal $\lambda$
with
\begin{align}
         V(\phi) + \frac{1}{4}\lambda\phi^4 \geq 0.
\end{align}
We also define $\Delta U$ and $\Delta \phi$ 
at the point where the equality holds:
\begin{align}
         \Delta U \equiv \frac{1}{4} \lambda_{\rm cr}\Delta\phi^4 = -V(\Delta \phi).
\end{align}
We can also define $\phi_c$ as the maximal value of $\phi$ with
\begin{align}
        V(\phi_c)=\frac{1}{4} \lambda_{\rm cr}\phi^3_c\Delta \phi.
\end{align}
Then, $V_{\rm max}$ is given by a maximal value of $V(\phi)$ in $\phi<\phi_c$.
As before, we define $r_*^2\equiv \Delta\phi^2/\Delta U$, $r_c
\equiv r_*(V(\phi_c)/\Delta U)^{-1/6}
$ and $m_{\rm typ}^2\equiv V_{\rm max}/\phi_c^2$. 
Then, if the condition Eq.~(\ref{eq:condmax}) does not hold,
bounce action will deviate from the lower bound.
Thus, this condition can be regarded as a necessary condition
for the bounce action to have a value close to the lower bound.

The condition Eq.~(\ref{eq:condmax}) characterizes a smallness
of the potential barrier.
This is because if $V_{\rm max}$ is small, $m_{\rm typ}$ also becomes small.
In addition, if $V_+$ is small, $r_c$ becomes large.
And if the barrier is relatively large, the bounce action will deviate from  
the lower bound $24/\lambda$.

\bibliography{ref}
\bibliographystyle{utphys}

\end{document}